# Exploiting Context-Awareness for Secure Spectrum Trading in Multi-hop Cognitive Cellular Networks

B. Lorenzo[*], I. Kovacevic[†], F. J. Gonzalez-Castano[*], J. C. Burguillo[*]

*AtlantTIC, University of Vigo, Spain

[†]Dept. Communication Engineering, University of Oulu, Finland

***Abstract*—In this paper, we consider context-awareness to enhance route reliability and robustness in multi-hop cognitive networks. A novel context-aware route discovery protocol is presented to enable secondary users to select the route according to their QoS requirements. The protocol facilitates adjacent relay selection under different criteria, such as shortest available path, route reliability and relay reputation. New routing and security-based metrics are defined to measure route robustness in spatial, frequency and temporal domains. Secure throughput, defined as the percentage of traffic not being intercepted in the network, is provided. The resources needed for trading are then obtained by jointly optimizing secure throughput and trading price. Simulation results show that when there is a traffic imbalance of factor 4 between the primary and secondary networks, 4 channels are needed to achieve 90% link reliability and 99% secure throughput in the secondary network. Besides, when relay reputation varies from 0.5 to 0.9, a 20% variation in the required resources is observed.**

## I. INTRODUCTION

Future 5G cellular networks face the challenge of acquiring adequate spectral resources to meet the increasing demand for new services and applications. Cognitive radio (CR) [1] emerged as a promising technology which endows wireless devices with the capability to adapt their operating parameters based on the radio environment and utilize the spectrum in an opportunistic manner. Moreover, enabling multi-hop communication can further increase bandwidth efficiency, coverage and provide ubiquitous connectivity. Despite the many advantages, multi-hop cognitive networks face unique security challenges.

Secondary users (SUs) have mostly used context-aware information enabled by CR for sensing purposes to detect the presence of primary users (PUs) [2] and also for physical layer security [3]. While context-awareness has been applied for routing decisions in ad-hoc networks [4], few works have considered it to make networking decisions in CR networks [5]-[7]. The application of reinforcement learning to dynamic channel selection, scheduling, and congestion control is addressed in [5], which shows it improves network performance in terms of throughput and stability, and significant reduces channel switching. In [6]-[7] the robustness of routing in CR is considered. These works focus on the coexistence of PUs and SUs, and provide solutions to secure PUs, but ignore security for SUs.

This work takes advantage of context-awareness to provide reliable routes and enable SUs to accurately estimate the resources for trading needed to satisfy their QoS requirements.

Spectrum trading for multi-hop secondary networks has attracted some attention lately [8]-[9]. These works ignore both the uncertainty of channel availability due to PU activity and QoS provisioning. Besides, they consider truthfulness in the bidding process and ignore the trustworthiness of the required resources. Thus, in order to prevent users from being treated as cheaters and lose the chance to obtain resources, additional mechanisms are needed to provide reliable routes.

In this paper, we exploit context-awareness to improve route reliability and robustness in multi-hop secondary networks. First, a novel context-aware route discovery protocol is developed, which is aware of both user availability to relay and available channels. The optimum number of channels and switching time are obtained to satisfy users' QoS requirements in terms of reliability and delay. Then, new routing and security-based metrics are defined to measure route robustness in the spatial, frequency and time domains. A usage-based pricing model is presented and the optimum number of resources to trade is obtained to satisfy the users' QoS requirements, and optimize secure throughput and price trade-off.

The main contributions of the paper are summarized as follows.

1) A network model for multi-hop cognitive cellular networks ($MC^2Ns$) that is context-aware and enables fine-grained adaptation to network requirements in a 3D scale (spatial, frequency and temporal domains). The model is mapped into an absorbing Markov chain to analyze a number of performance metrics (such as delay and reliability).

2) A novel context-aware route discovery protocol that considers the availability probability of the SUs to relay and the available channels according to PU activity. The protocol considers different relaying priorities such as shortest available distance, route reliability and relay reputation.

3) New routing and security-based metrics are defined to measure robustness on the spatial, frequency and temporal domains, and secure throughput.

The rest of this paper is organized as follows. The system model is described in Section II. In Section III, the context-aware route discovery protocol is presented and analyzed. Numerical results are given in Section IV. Finally, Section V concludes the paper.

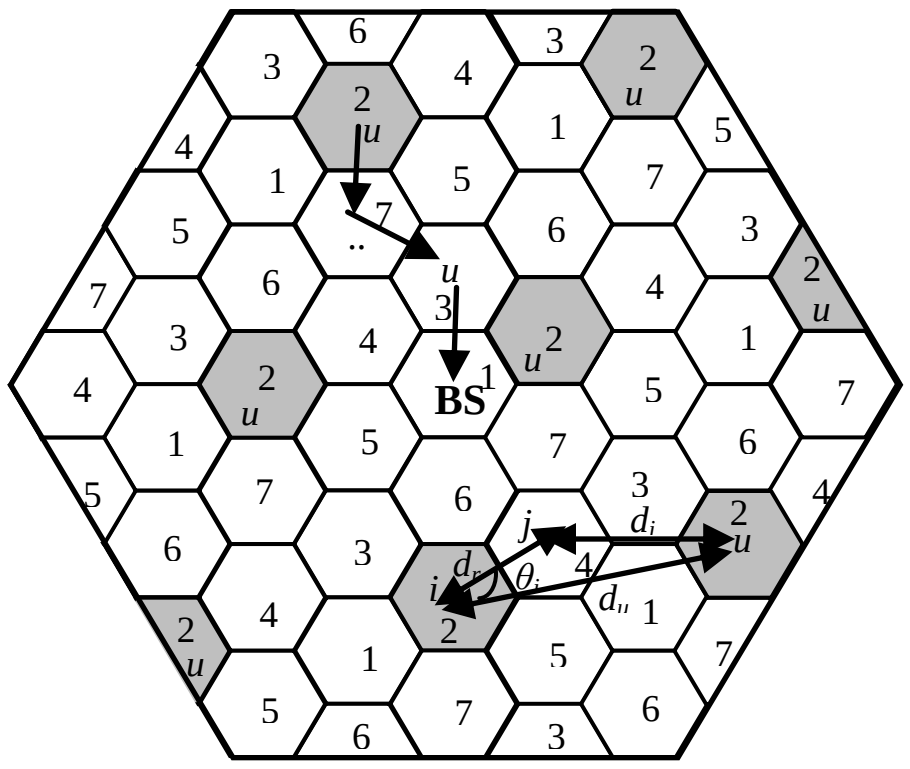

Fig. 1. Network model with $K$-reuse pattern for $K$ = 7.

## II. System Model

### A. Network Model

We consider a densely deployed multi-hop cognitive cellular network (MC$^2$N) as shown in Fig.1. For modeling purposes, the area of the cell is divided into hexagonal subcells[1] of radius $r$. In each subcell, there is, potentially, a SU that may act as a source, relay or destination. We assume that a spectrum owner or primary operator (PO) leases its idle channels to SUs. Due to PU activity, there is uncertainty regarding channel availability (PU activity is modeled in detail in Section III.C). Each SU is equipped with one radio capable of switching between different channels. The SU source will transmit to its intended destination by relaying to adjacent SUs (located in adjacent subcells) through the available channels. Besides, an adjacent SU will be available to relay with probability $p$, which depends on coverage, battery charge, and willingness to cooperate.

### B. Tessellation Factor and Scheduling

We assume that the BS is surrounded by $H$ concentric rings of subcells. In the example in Fig.1, $H$=3. For simplicity, we approximate that in each subcell the SU potentially available to relay is located in the center of the subcell. The relaying distance between adjacent subcells is denoted by $d_r$ and is related to the subcell radius as $d_r = \sqrt{3} \cdot r$.

We assume that a user $i$ can successfully transmit to its adjacent relay $j$ when the received power at $j$ exceeds the receiver sensitivity $\varepsilon$. For a given relaying distance $d_r$, the minimum transmission power for user $i$ is $P_{i,min} = \varepsilon \cdot (d_r)^{\alpha}$, where $\alpha$ is the path loss factor. Users are interested in transmitting with the minimum power possible $P_i = P_{i,min}$ to reduce interference and power consumption. For simplicity, we assume that the tessellation factor $r$ is the same for all subcells and thus, users transmit with the same power $P_i = P$.

Let us assume that user $i$ transmits to its adjacent user $j$ and at the same time there is an interfering user $u$ transmitting on the same channel as shown in Fig. 1. Then, the signal to interference plus noise ratio at any relay is given by [10]

[1]This partition is not physically implemented in the network, but rather used to capture the mutual relations between the terminals in the cell that are potentially available to relay each other's messages [10].

$$SINR(w_i,P) = \frac{PG_{ij}^{w_i}}{\sum_{u=1}^{n_{wi}} PG_{uj}^{w_i} + N_r} = \left( \frac{N_r d_r^{\alpha}}{P} + \sum_{i=1}^{n_{wi}} \left( \sqrt{\frac{1}{1+K-2\sqrt{K}\cos\theta_i}} \right)^{\alpha} \right)^{-1} \quad (1)$$

where $G_{ij}^{w_i}$ is the channel gain between user $i$ and $j$ on channel $w_i$, $n_{wi}$ is the number of concurrent transmissions using that channel, $G_{uj}^{w_i}$ is the channel gain between interfering users $u$ and $j$ on channel $w_i$, $N_r$ is the background noise power, $\theta_i$ is the separation angle between user $i$ and receiver $j$ and interfering user $u$ and, $K$ is the tessellation factor explained below. As we are considering a dense network, the channel model adopted includes propagation losses, but not the effects of fading due to the proximity between the users [10]. The Shannon capacity on link $l$ is given by $c_l = \log(1+SINR(w_i,P))$ and the capacity on the route $R$ is $c_R = \min_R c_l$.

Thus, the normalized route capacity when $w_R$ channels are used on the route can be obtained as

$$\overline{c}_R = c_R / w_R. \quad (2)$$

The optimization of the scheduling in a multi-hop network is an NP-hard problem [11], [12]. To keep the scheduling process simple, we apply a conventional resource reuse scheme for cellular networks to our tessellation scheme, shown in Fig. 1, for the resource reuse factor $K$ = 7. The subcell index within the cluster indicates the slot allocation, $k$ =1,..., $K$. Users with the same slot index will transmit simultaneously. The transmission turn (in a round robin fashion) is given by the slot index. To avoid transmission/reception collisions, users can transmit simultaneously if they are separated by a distance $d > 2d_r$. This constraint is a direct consequence of the fact that users are equipped with a single radio. In this way, transmissions will be collision-free regardless of the adjacent relay or channel chosen for transmission. This constraint holds for $K \geq 7$. To take advantage of all available channels and reduce the interference level in the network, we will randomly assign a different channel to each subcell that shares the same transmission slot.

The $K$-scheduling pattern proposed is a heuristic that avoids the complexity of re-computing the schedule according to PU activity and user availability to relay.

### C. Cognitive Channel Availability

We model channel availability when a routing/relaying decision is made. For this purpose, we consider data session arrivals into a cell of a cellular network with capacity of $c$ channels and determine the probability that the primary operator will not use $b$ out of $c$ channels. By modeling arrivals as a Poisson process and session duration as exponentially distributed, this probability can be obtained as a solution of the birth/death equations for a conventional M/M/c system. This model has been widely used in the literature [13]. If we consider that the arrival rate is $\lambda_{Pn} = \lambda_P$ for $n$ PUs and the service rate is $\mu_P$ (they are state-independent), then the probability that the PO will not use $b$ out of $c$ channels is [14]

$$p_b = \frac{\lambda_P^{c-b}}{(c-b)!\,\mu_P^{c-b}} p_0, \quad (1 \leq b < c) \quad (3)$$

where $p_0$ is obtained for $r_P = \lambda_P / \mu_P$ and $\rho_P = r_P / c < 1$ as

$$p_0 = \left( \sum_{n=0}^{c-1} \frac{r_P^n}{n!} + \frac{r_P^c}{c!(1-\rho_P)} \right)^{-1}$$

The SU will have a channel available as long as there is at least one channel that is not being used by the PO. Hence, the probability that there will be at least $a$ channels available is

$$p_a = \sum\nolimits_{b=a}^{c-1} p_b \tag{4}$$

where $p_b$ is obtained using (3).

We assume that spectrum monitoring is perfect and focus on the probability that a PU returns to a channel currently allocated to an SU. This is checked at every hop of the SU route. We approximate the calculation by assuming that the average service time of the SU is $1/\mu_S$. Then, the probability of $k_P$ new PUs arriving within that time interval is [14]

$$p_k(t = 1/\mu_S) = \frac{(\lambda_P t)^{k_P}}{k_P!} e^{-\lambda_P t} = \frac{(\lambda_P / \mu_S)^{k_P}}{k_P!} e^{-\lambda_P / \mu_S}$$

The probability that a specific channel out of $b$ is allocated to one of the $k_P$ new arrivals is $k_P/b$. So, the average probability of PU return to a particular channel is

$$p_{b,return} = \sum_{k_P=0}^{b} \frac{k_P}{b} p_k(t = 1/\mu_S) = \sum_{k_P=0}^{b} \frac{k_P}{b} \frac{(\lambda_P / \mu_S)^{k_P}}{k_P!} e^{-\lambda_P / \mu_S} \tag{5}$$

If a PU returns to the channel currently allocated to a SU, the transmission will be interrupted and the SU will be forced to try a new option. Equation (5) can be further averaged out to give the average PU return probability defined as

$$p_{return} = \sum\nolimits_b p_b p_{b,return} \tag{6}$$

where $p_b$ is given by (3).

The impact of PU activity on route discovery is considered in the following section.

## III. Context-Aware Route Discovery Protocol

This section presents a route discovery protocol for SUs and its analysis. By following the $K$-scheduling pattern, collisions between adjacent relays are avoided and thus, we focus on discovering the route based on users' availability to relay and PU activity.

### A. *Description of the Protocol*

Let us assume that in subcell $i$ there is an SU willing to transmit to an intended destination by relaying to adjacent users. As each user has 6 adjacent subcells, the candidate relay may be in any of these subcells $m$, $m$ = 1,.., 6. We start by assuming that the route discovery protocol provides the shortest available path. Then, according to the relay priorities given by the distance to the destination, the SU first checks the availability of the adjacent user (located in the adjacent subcell) in the direction corresponding to *the shortest distance* towards the destination. The user will be available to relay with probability $p$. If so, the SU will transmit to the adjacent subcell by using a cognitive channel, and this transition will occur with probability $p_{m=1}$. If the relay in this adjacent cell is not available, the SU will check the availability of the next adjacent user (2nd user), as shown in Fig. 2. If that user is available, the SU will relay to him/her. This transition will occur with probability $p_{m=2}$. Otherwise, the protocol continues in the same fashion until it checks the last adjacent user (6th user). If in the last adjacent subcell there is no user available to relay, the route will not be established, which happens with probability $p_0$. This protocol will be referred to as *shortest available path routing* (SaPR).

As channel availability depends on the activity of the PUs, in any system state a PU may return to the channel after the transmission to a relay is initiated. This is illustrated in Fig. 2 as well. In this case, the process will be aborted (with probability $p_{return}$) and the SU will try another channel.

The relaying probability from subcell $i$ to any adjacent subcell $j$ can be obtained by mapping the transmission pair $(i,j) \rightarrow m$, as

$$p_m = p_a p(1-p)^{m-1} \xi, \; m = 1,\ldots,6 \tag{7}$$

where $p_a$ is the probability that the SU has a channel available and is obtained by (4) when $p_a(a = 1)$, $p$ is the probability that in the adjacent subcell there is a user willing to relay, and $\xi$ is the probability that there will not be a PU return to the channel, $\xi = 1 - p_{return}$, and $p_{return}$ is given by (6). Parameter $\xi$ will be used in Section III.C to measure link reliability.

The overall probability of relaying to any adjacent subcell is

$$p_t = \sum\nolimits_m p_m \tag{8}$$

where $p_m$ is given by (7). The probability that the SU will not be able to transmit to any of the adjacent users is $p_0 = 1 - p_t$, which is represented as a transition to an absorbing state *nr* (*no route*).

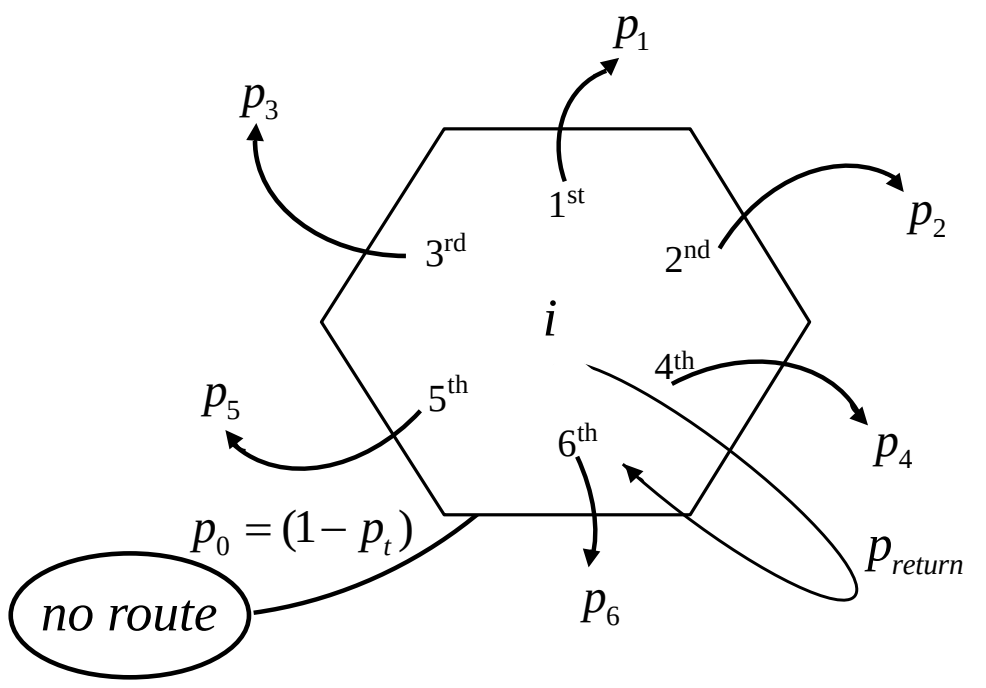

Fig. 2. Representation of the state transition probabilities given by (7).

### B. *Analysis of the Protocol*

We define a relaying probability matrix $\mathbf{P} = [p_{ij}]$ where entry $p_{ij}$ indicates the relaying probability from subcell $i$ to $j$. This probability can be obtained from (7) by mapping the transmission from subcell $i$ to any adjacent subcell $j$ as $(i,j) \rightarrow m$. It is worth noticing that $\mathbf{P}$ also defines the network topology. To analyze the relaying process in the network, we map the tessellation scheme into an absorbing Markov chain with a set of absorbing states $\mathcal{A}_s = \{D, nr\}$. These states represent the end of the route when the user has reached the destination (e.g., BS, AP, mobile to mobile connection) or when no route ($nr$) is available.

Then, we reorganize the relaying matrix into an $(N+1)\times(N+1)$ matrix of the form [15]

$$\mathbf{P}^* = \begin{bmatrix} \mathbf{I} & \mathbf{0} \\ \mathbf{R} & \mathbf{Q} \end{bmatrix} \quad (9)$$

where $N$ is the number of subcells, $\mathbf{I}$ is an $N_A \times N_A$ diagonal unitary matrix corresponding to the number of absorbing states, $\mathbf{0}$ is an $N_A \times (N - N_A + 1)$ all-zero matrix, $\mathbf{R}$ is the $(N - N_A + 1) \times N_A$ matrix of transition probabilities from transient states to absorbing states and $\mathbf{Q}$ is the $(N - N_A + 1)\times (N - N_A + 1)$ matrix of transition probabilities between transient states.

By defining $\mathbf{N} = (\mathbf{I} - \mathbf{Q})^{-1}$, the mean time for the process to reach any absorbing state starting from transient state $i$ is [15]

$$\boldsymbol{\tau} = (\tau_1, ..., \tau_{N-N_A+1})^t = T(\mathbf{I}-\mathbf{Q})^{-1}\mathbf{1} = T\mathbf{N}\mathbf{1} \quad (10)$$

when the dwell time for any state $i$ is the same, $T = T_i$ and $\mathbf{1}$ is an $(N–N_A+1)\times 1$ column vector of ones. Otherwise, $\boldsymbol{\tau}=(\mathbf{I}–\mathbf{Q})^{-1}\boldsymbol{v} = \mathbf{N}\boldsymbol{v}$ where $\boldsymbol{v}$ is a column vector whose components are $T_i$.

For the normalized dwell time $T = T_i = 1$, the entries $\tau_i$ of vector $\boldsymbol{\tau}$ represent the average number of hops (route length) from state $i$ to the absorbing state.

The probability that the Markov process starting in a transient state $i$ will end up in an absorbing state $j$ is $e_{ij}$, where

$$\mathbf{E} = [e_{ij}] = (\mathbf{I} - \mathbf{Q})^{-1}\mathbf{R} \quad (11)$$

If we denote by $\mathbf{f}$ the vector of probabilities of initial user positions, the probabilities of accessing the destination and finding no route are

$$[\mathbf{p}_D, \mathbf{p}_{nr}] = \mathbf{f}\mathbf{E} \quad (12)$$

where, $\mathbf{p}_D = (p_{D,1}, ..., p_{D,N-N_A+1})^t$ and $\mathbf{p}_{nr} = (p_{nr,1}, ..., p_{nr,N-N_A+1})^t$.

### C. Route Reliability

We define route reliability as the probability of non PU return on the route, $\xi_R = \prod_{l\in R} \xi_l$, where $\xi_l$ is the link reliability on every hop of the route $R$. In order to determine the resources needed to guarantee a certain level of route reliability, we will start by analyzing the reliability per link.

Let us assume that the duration of a single hop transmission is given by the SU service rate as $t_S = 1/\mu_S$. To avoid collision of the incoming PU on the channel being used by an SU, the SU will switch to another channel before this happens.

We consider that PU arrivals on every hop are independent and identically distributed (i.i.d). Hence, link reliability is the same on every hop of the route, $\xi_l = \xi$. Therefore, route reliability can be approximated as $\xi_R = \xi^\tau$, where $\tau$ is the route length given by (10). As a reminder, link reliability, $\xi = 1 – p_{return}$, decreases in time as the probability of PU return exponentially increases. For a minimum required level of route reliability, $\xi_{R,min}$, the equivalent minimum link reliability is $\xi_{\min} = \sqrt[\tau]{\xi_{R,\min}}$. As the number of channels are limited, the switching interval should be the maximum that satisfies the link reliability requirements. Thus, the channel switching time per link can be determined as

$$t_w^* = \operatorname{argmax}_{t_w}\ t_w \cdot (\xi(t_w) - \xi_{\min}) \quad (13)$$

where $t_w^*$ is the maximum duration of the time interval that satisfies the link reliability $\xi_{min}$. The more restrictive $\xi_{min}$ is, the more often the channel should be switched to avoid PU return.

The number of times the channel is switched per link per message is thus, $n_w = t_S / t_w$. As PU and SU arrivals are independent, the minimum number of backup channels needed to satisfy the reliability requirement per hop can be determined as follows,

$$w_{\min} = \begin{cases} n_w, & t_P \geq t_S \\ n_w t_P / t_S, & t_P < t_S \end{cases} \quad (14)$$

which depends on the PU service time $t_P$, $t_P = 1/\mu_P$. If $t_P \geq t_S$, $n_w$ different channels are needed per link; otherwise the channels will be reused with a reusability factor $t_P / t_S$.

In general, link reliability can be increased by considering a number of backup channels $w$ and switching the channel at intervals $t = t_w = t_S / w$. The new $p_{return}$ can be obtained by (5)-(6) for the previous time interval. Besides, the relaying probability should be modified as

$$p_m = p_a(a = w)p(1-p)^{m-1}\xi(t_w)\,,\ m = 1,\ldots,6 \quad (15)$$

where $p_a$ is the probability that the SU has $w$ channels available and is obtained by (4) when $p_a(a = w)$, $p$ is the user willingness to relay, and $\xi$ is the probability that there will not be a PU return to the channel, $\xi = 1 - p_{return}$, with $p_{return}$ in interval $t = t_w$ given by (6). Equations (9)-(12) can be obtained as before for the new relaying probability.

Users also have QoS requirements in terms of delay. For a maximum tolerable delay, $\tau_{\max}$, the number of backup channels $w$ needed is

$$w^* = \operatorname{argmin}_{w \geq w_{\min}}\ (\tau_{\max} - \tau(p, w))^2 \quad (16)$$

where $\tau$ is given by (10), $p$ is the relay availability probability, and $w_{min}$ is obtained in (14). It is worth noting that the number of backup channels $w$ will impact the delay $\tau$ as the relaying probability will be affected in (15).

Channel reusability between adjacent hops depends on the relation between $t_P$ and $t_S$. Based on this, the minimum number of channels the SU needs to buy from the PO to satisfy QoS requirements in terms of reliability and delay is

$$w_{R,\min}^* = \begin{cases} w^*, & t_P \leq t_S \\ w^* t_P / t_S, & t_P > t_S \end{cases} \quad (17)$$

where $w^*$ is the optimum number of backup channels per hop. If $t_P \leq t_S$, the same channels can be reused in adjacent hops. Otherwise, the number of channels needed per route increases by factor $t_P / t_S$.

### D. Security Improvements

#### D1. Relay Reputation

We obtain the reputation of the relays based on the previous experiences of users with a relay in terms of packet forwarding. The reputation varies from 0 to 1, where 0 is completely untrustworthy and 1 is the opposite. We denote by $s_{ij}^a$ the advertised security level by a relay $j$. When user $i$ relays to $j$, it will calculate the relaying experience $s_{ij}^e$ as $s_{ij}^e = p_t$, where $p_t$ is the probability that the user will transmit to an adjacent relay (8). Combining both factors, the current experience $s_{ij}^c$ is obtained as $s_{ij}^c = s_{ij}^e / s_{ij}^a$ . If we consider past trustworthiness histories, the probability that user $i$ will have a safe connection when relaying to $j$, $s_{ij}$, can be calculated using the exponential moving average as

$$s_{ij}(t) = \begin{cases} \alpha_e s_{ij}^c(t) + (1-\alpha_e) s_{ij}(t-1), & \text{if } t > 1 \\ s_{ij}^c(t), & \text{otherwise} \end{cases} \quad (18)$$

where $\alpha_e$ is a weighting factor between 0 and 1 and $t$ represents the current episode within a particular window of episodes. Large $\alpha_e$ values will assign more weight to current episodes than previous ones.

The context-aware route discovery protocol can be modified to include relay reputation by introducing the level of trust into the relaying probability in (7) as

$$p_m = p_a p (1-p)^{m-1} \xi s_m, \; m = 1,\dots,6$$

where $s_m$ is obtained by mapping the transmission pair as $(i,j) \rightarrow m$. The resulting routing protocol will be referred to as *secure path routing* (SecPR).

*D2. Probability of being intercepted*

We measure the probability that an SU transmission will be intercepted in spatial, frequency and time domains in our MC$^2$N model. We denote by $\sigma_j$ the robustness to be intercepted in dimension $j$, $j \in \{s, f, t\}$. The SU will be intercepted if it is intercepted in all three dimensions. Thus, the eavesdropper needs to know the subcell from which the user is transmitting as well as the frequency and time slot. In a route of $\tau$ hops, the probability of interception is

$$p_d = 1 - \left[ 1 - \prod\nolimits_{j \in \{s,f,t\}} (1-\sigma_j) \right]^{\tau} \quad (19)$$

which is the probability that the transmission will be intercepted in at least one hop of the route.

*Spatial robustness*

Let assume that there are $k$ eavesdroppers randomly located in the network. The probability that the SU will be intercepted in space is $k/N$, where $N$ is the number of subcells. Spatial robustness is thus defined as

$$\sigma_s = 1 - k / N. \quad (20)$$

*Frequency robustness*

We assume that an eavesdropper can observe all available channels $c$ in the macrocell. If we consider that the eavesdropper selects one channel and waits to detect the transmission on that channel, then frequency robustness is

$$\sigma_f = 1 - 1/c. \quad (21)$$

*Time robustness*

We denote by $t_o$ the eavesdropper observation time per day, which is the amount of time the eavesdropper is actively observing the network, and $t_S$ time duration of the SU transmission. If we assume that $t_o > t_S$, time robustness is

$$\sigma_t = 1 - t_o / 24. \quad (22)$$

*E. Throughput and Pricing*

We define the normalized throughput as $\overline{T} = \overline{c}_R / K\tau$, where $\overline{c}_R$ is the normalized capacity when $w_R$ channels are used in the route and is given by (2), $\tau$ is the route length as in (10), and $K$ is the duration of the scheduling. Then, a new metric called normalized secure throughput is defined as the portion of traffic not being intercepted,

$$\overline{T}_s = p_{nd} \overline{T} \quad (23)$$

where $p_{nd}$ is the probability of non-interception, $p_{nd} = 1 - p_d$.

We adopt a usage-based pricing scheme, where the primary operator charges the SU for the number of channels requested and the duration of the transmission. We assume a fixed fee per channel $\Phi_w$ and a time unit of transmission $\Phi_t$. Then, the pricing function is represented as follows,

$$price = w_R \Phi_w + \tau t_s \Phi_t \quad (24)$$

where $w_R$ is the number of channels requested and $\tau \cdot t_s$ is the duration of the transmission, with $\tau$ representing route length and $t_s$ service time.

Then, if we define the utility of the SU as $U = \overline{T} - \gamma \cdot price$, the optimum number of resources $w_R$ needed to satisfy the reliability and delay requirements can be obtained as

$$\begin{aligned} &\underset{w_R}{\text{maximize}} \quad U = \overline{T} - \gamma \cdot price \\ &\text{subject to} \quad w_R \geq w_{R,\min} \end{aligned} \quad (25)$$

where $\gamma$ is a scaling parameter and $w_{R,min}$ is the minimum number of channels needed to satisfy the QoS requirements as obtained in (17).

## IV. Simulation Results

We conducted Monte Carlo simulations in Matlab to evaluate network performance with our context-aware solutions. The results were obtained over 1000 realizations. We considered the network shown in Fig. 1, where the radius of the macrocell is $R = 1000$ $m$, $H = 4$, and $P = 0.75$. The path loss exponent is $\alpha = 2$ and the noise power spectral density is $N_r = 10^{-4}$ W/Hz. We assume that there are $c = 10$ channels in the cellular network, $n$ of which are occupied by PUs, and $a$ potentially available channels for SUs. The available channels during a given transmission slot are randomly allocated to the SUs sharing that slot, according to the $K$-reuse pattern.

### A. Route Reliability

In Fig. 3, we present the utility defined as in (13) versus the switching time $t_w$ for different traffic situations in the secondary network. We can see that for a fixed $\lambda_P$ and a target reliability requirement $\xi_{min}$, the optimum switching time $t_w^*$ increases with the secondary user service rate $\mu_S$, as the probability of a return during the secondary user service time is lower.

In Fig. 4, the minimum number of channels $w_{min}$ to satisfy a given $\xi_{min}$ is shown. We assume $\mu_P = 4\lambda_P$ and different secondary user service rates $\mu_S$. We can see that $w_{min}$ is lower when networks are more imbalanced (i.e., $\mu_S = 8\lambda_P$). A reliability requirement of $\xi_{min} = 0.90$ can be achieved with 2 to 4 channels when $\mu_S = 8\lambda_P$ to $4\lambda_P$. However, when $\mu_S = 2\lambda_P$, the number of channels is doubled. In all cases, if the reliability requirement increases from 0.9 to 0.95, $w_{min}$ increases significantly.

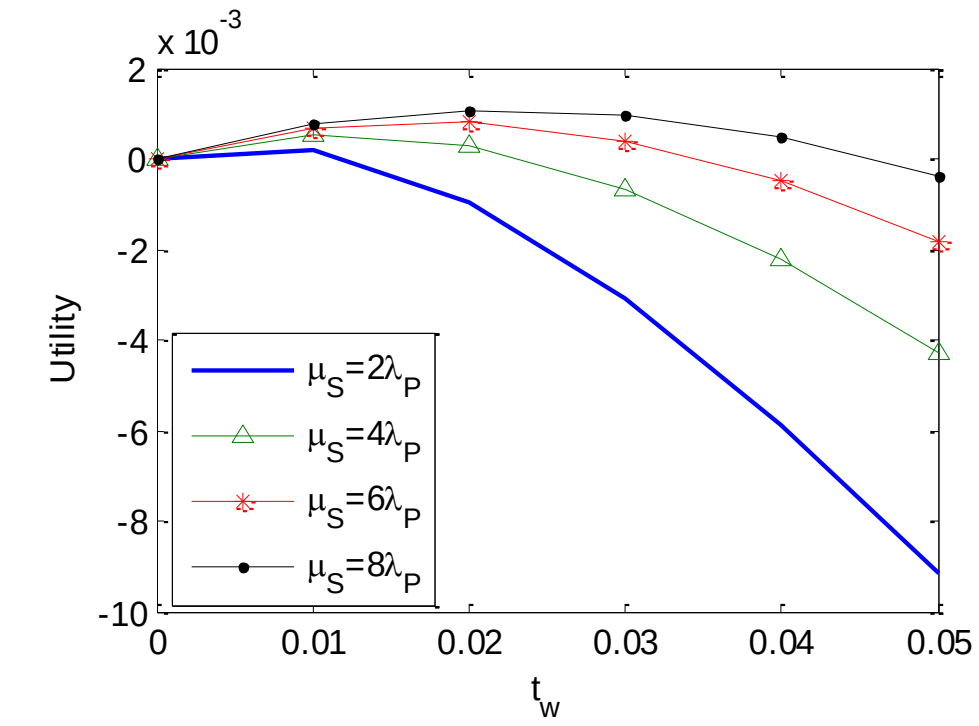

Fig. 3. Utility (13) versus $t_w$ for $\xi_{min}$ = 0.9.

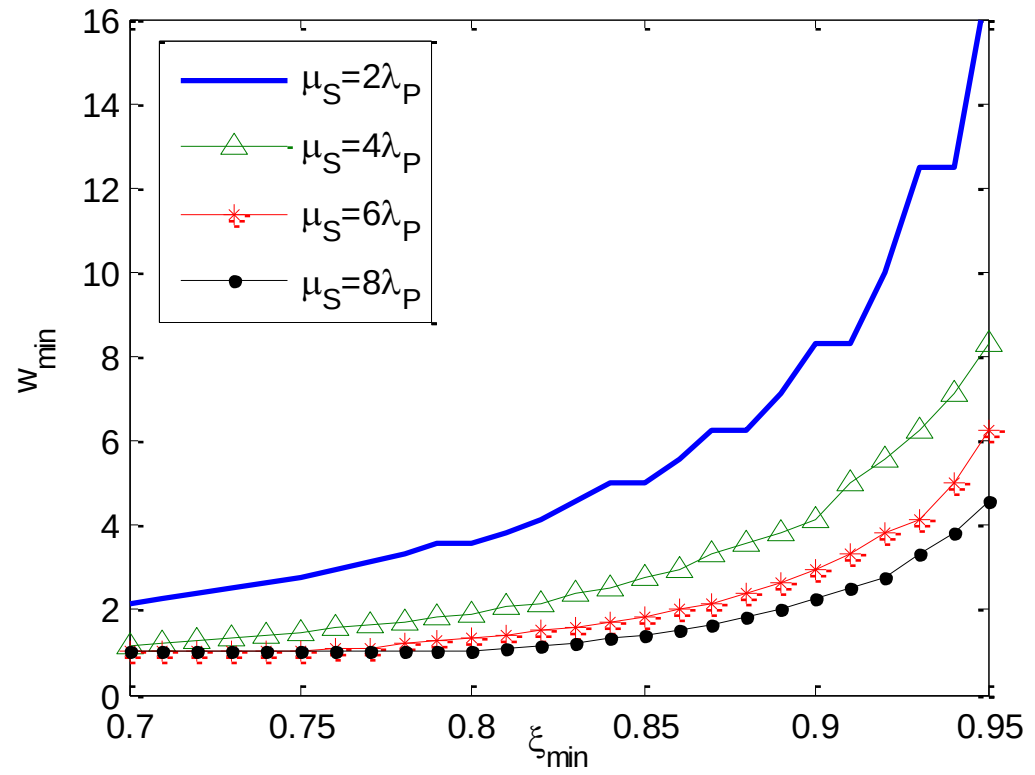

Fig. 4. Minimum number of channels $w_{min}$ versus $\xi_{min}$.

In Fig. 5, the utility defined in (16) is shown versus $w$ for the previous traffic situations. The relay availability probability $p$ was fixed to 1 and the reliability requirement to $\xi_{min}$ = 0.9. A maximum tolerable delay of $\tau_{max}$ = 3.5 was assumed. We can see that the previous QoS requirements are achieved with $w^*$ = 2 when $\mu_S = 8\lambda_P$, with $w^*$ = 3 when $\mu_S = 6\lambda_P$, with $w^*$ = 4 when $\mu_S = 4\lambda_P$ and $w^*$ = 8, otherwise. It is worth noting that to satisfy this QoS requirement, the optimum number of channels $w^* = w_{min}$. In the rest of the simulations $\mu_S = \mu_P = 4\lambda_P$ was considered and thus, $w_R^* = w^*$.

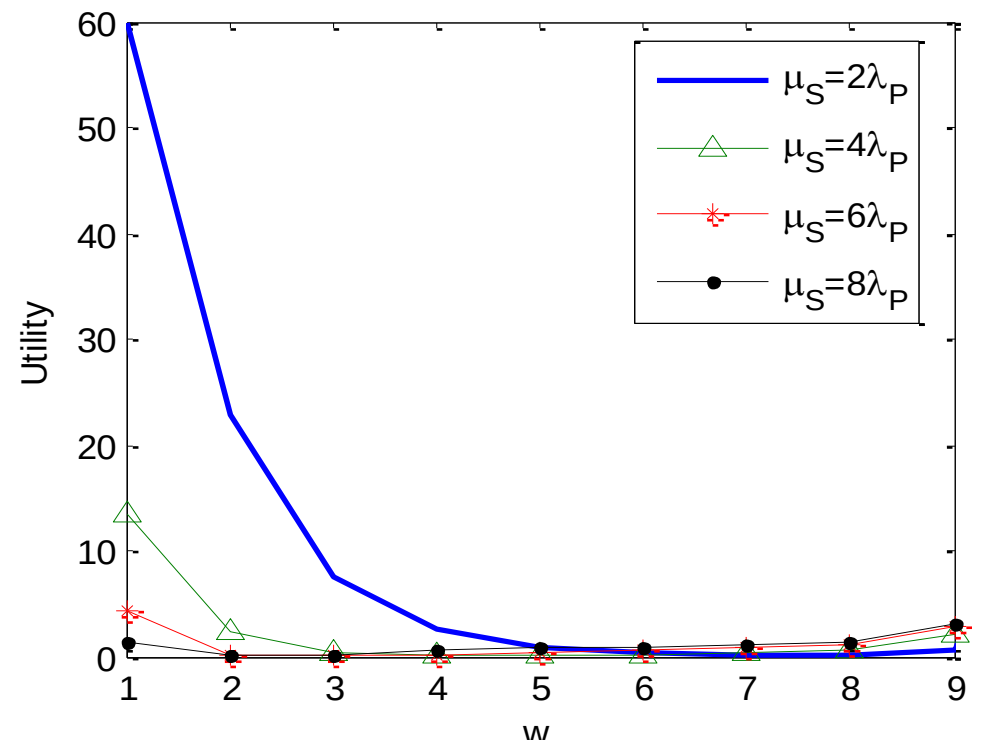

Fig. 5. Utility (16) versus $w$ for $\xi_{min}$ = 0.9, and $\tau_{max}$ = 3.5 when $p$ = 1.

## B. Security Improvements

We considered a relay advertising a security level of 0.9. Based on the previous experience a user had with this relay, the advertised security level was only provided for 15% of the time. For the remaining time, a security level between 0.5 and 0.9 was obtained. The relay reputation is shown in Fig. 6 for different observation instants $t$. The results were obtained for $T$ = 100 observations. We can see that, for a weight $\alpha_e = 2/(T+1) = 0.02$, the algorithm converged to a reputation $s$ = 0.75. However, if $\alpha_e$ increased, the relay reputation oscillated and convergence was slower or did not occur.

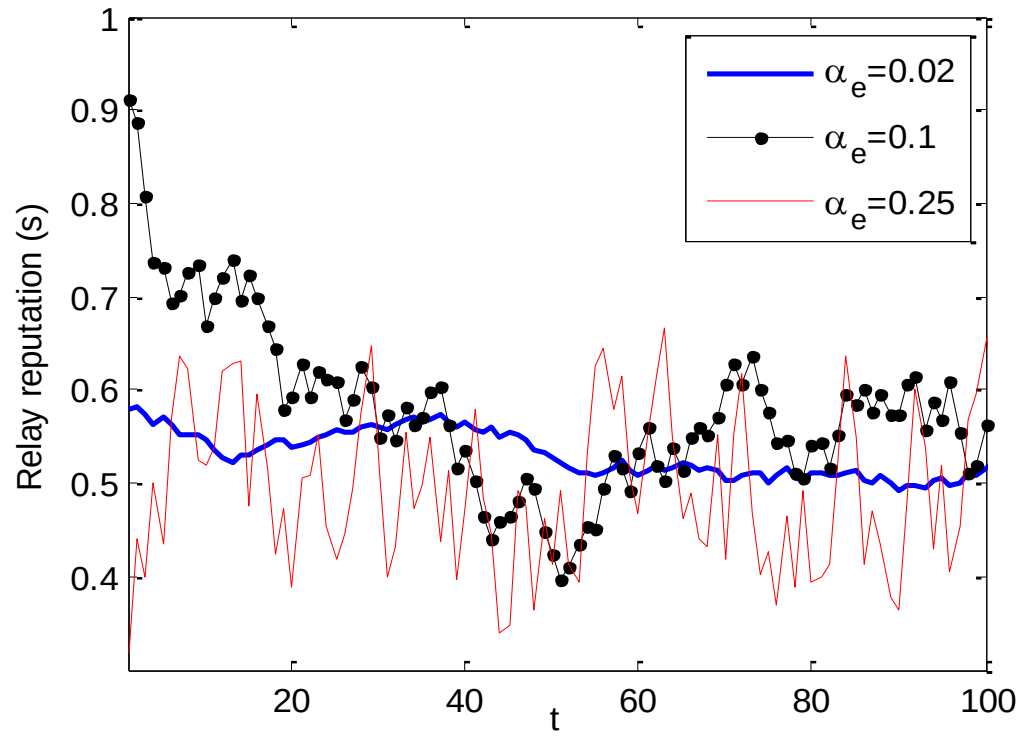

Fig. 6. Relay reputation ($s$) as in (18) versus the observation ($t$).

In Fig. 7 we present the average delay versus the weight $\alpha_e$ used to estimate the reputation of the relay in (18). It is worth noting that when $\alpha_e < 0.5$ (the past history has more weight than the current observation), the delay oscillated less. For small values of $\alpha_e$ ($\alpha_e$ = 0.02) the delay was almost constant. When we only considered the current observation, as in existing routing protocols (e.g., [16], [17]), the delay could incur in an error of $\pm$10% on the final delay. It can be also observed that the delay decreased for higher $p$.

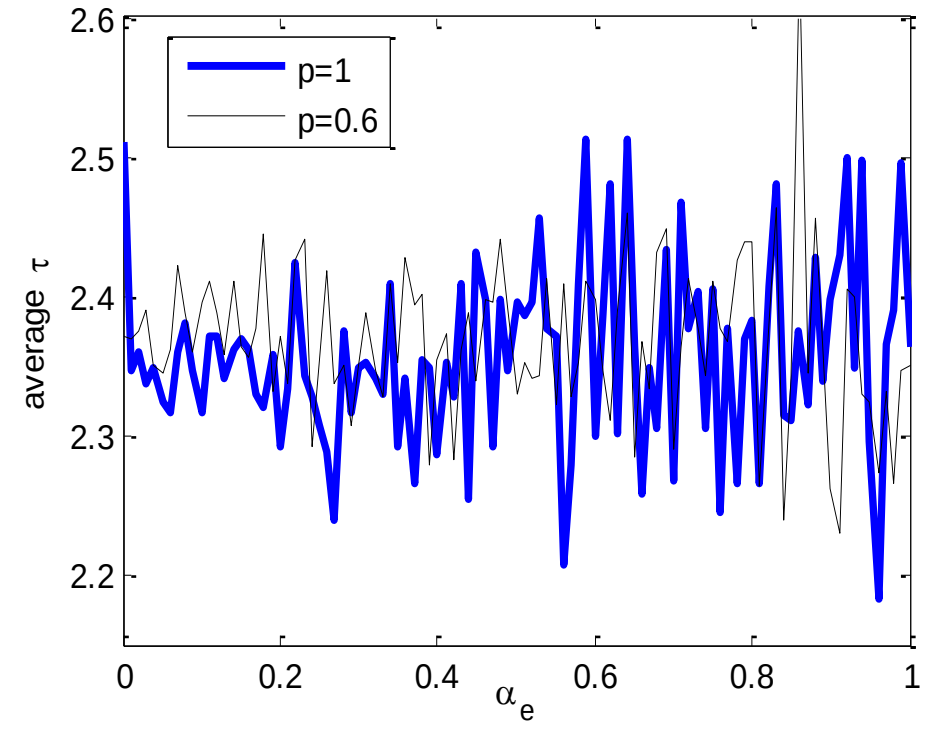

Fig. 7. Average delay considering relay reputation as in (18) versus $\alpha_e$.

In Fig. 8, the probability of not being intercepted, $p_{nd}$, is shown versus $w_{min}$ for different availability probabilities $p$. We assumed that there was an eavesdropper in 10% of the subcells and that the observation time was 8 hours per day. We can see that $p_{nd}$ increases with $p$ and $w_{min}$ as the delay $\tau$ is lower. Every additional 10% increase in the number of eavesdroppers resulted in an additional 1% decrease in $p_{nd}$. These results are not shown due to space constraints.

In Fig. 9, the throughput and secure throughput are shown versus $p$ for different channel requirements, $w_{R,min}$. We can see that 99% of the traffic was safely transmitted in the 3D domain. If $w_{R,min}$ increased by 2 channels, the throughput decreased by about 15% in average. Besides, as the required $w_{R,min}$ increased, the effects of the availability probability on throughput become negligible.

In Fig. 10, the utility defined as in (25) is shown versus $w_{R,min}$. We considered $\tau_{max}$ = 3.5. The $w_{R,min}$ values needed to satisfy this requirement are shown along the dash line. We can see that when $p$ = 0.6, the maximum utility was obtained for $w_{R,min}$ = 3 ($\xi_{min}$ = 0.87). If the user requires higher reliability, the utility decreased as the price was higher. For higher $p$ the optimum utility was obtained for $w_{R,min}$ = 2 ($\xi_{min}$ = 0.80).

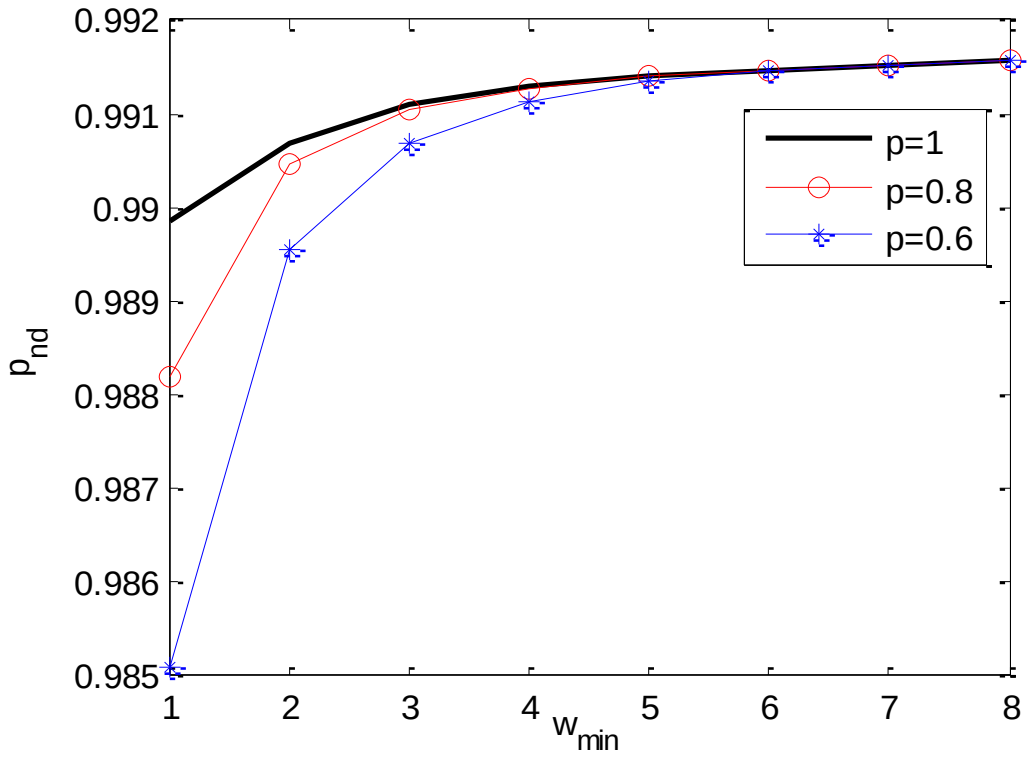

Fig. 8. Probability of avoiding interception versus $p$.

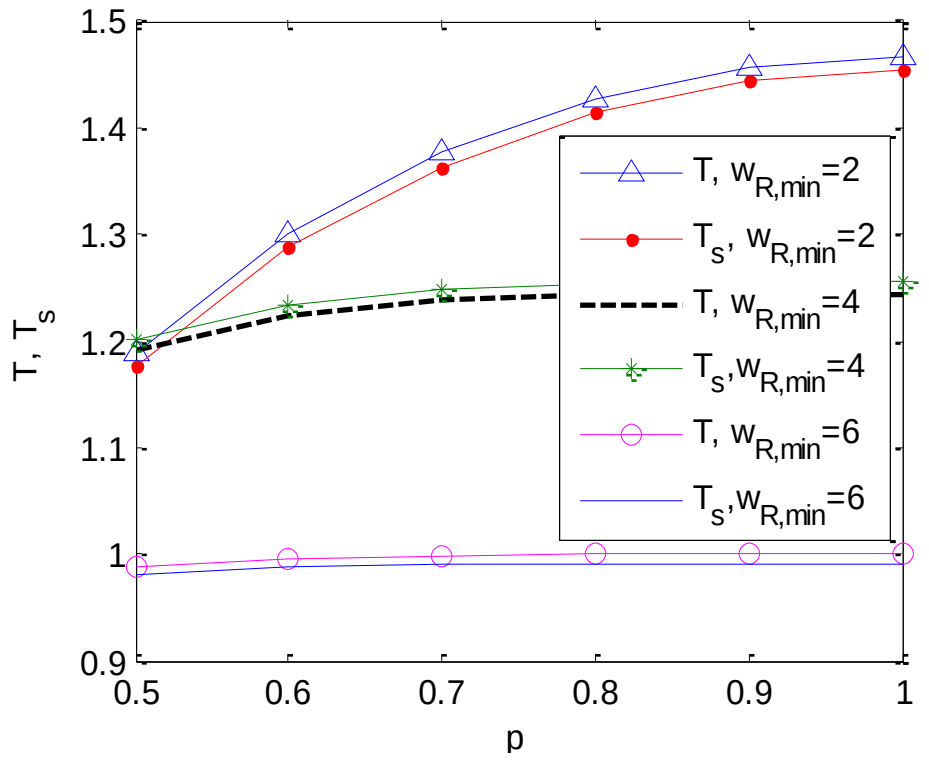

Fig. 9. Throughput versus $p$.

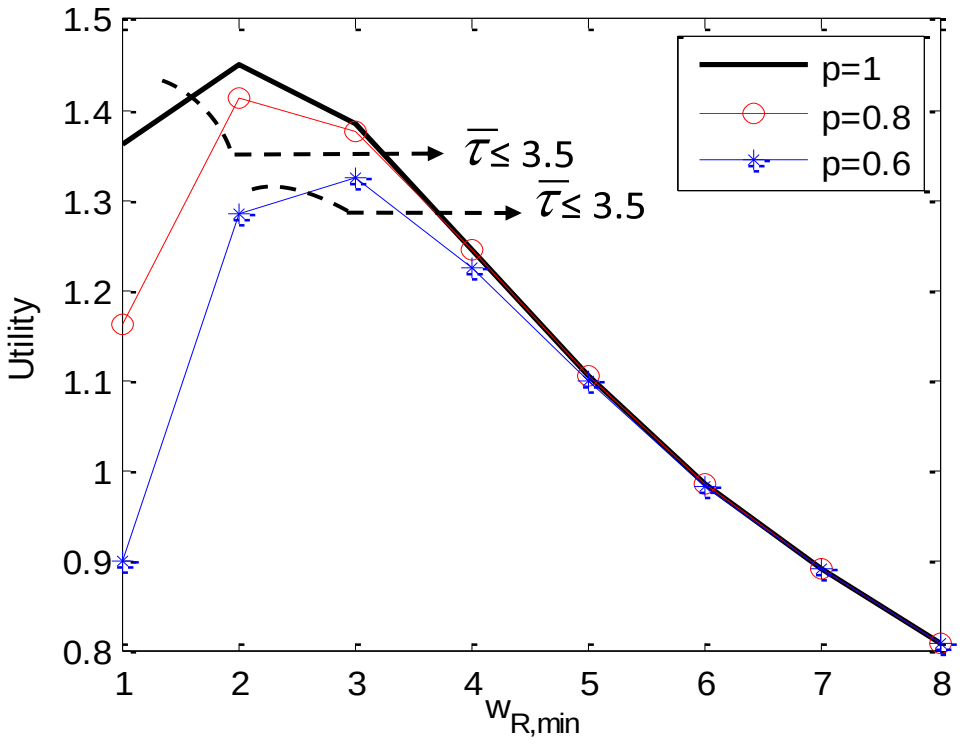

Fig. 10. Utility (25) versus $w_{R,min}$.

## V. CONCLUSIONS

In this paper, we consider context-awareness to increase route reliability and robustness in multi-hop cognitive cellular networks. A context-aware route discovery protocol that uses different criteria for route discovery (e.g., shortest available path, safest path) is presented. User QoS requirements are considered in route selection. Optimum switching time and number of channels are obtained to guarantee the required level of link reliability and delay. Then, new routing and security-based metrics are defined to measure robustness in space, time and frequency domains. A new metric denoted as security throughput is also defined to indicate the percentage of traffic not being intercepted in the network.

The results of extensive simulations conducted show that the reliability of the secondary network mainly depends on traffic imbalance between the secondary and primary networks. When there is an imbalance of factor 4, 4 channels are needed per link to guarantee a link reliability of 90%. Besides, robustness to interception in the spatial, frequency and time domains results in a secure throughput of 99%. For higher imbalances in both networks, even larger improvements are expected. Moreover, when relay reputation varies from 0.5 to 0.9, a 20% variation in resources is observed. The results are useful for determining the resources needed for secure spectrum trading.

## ACKNOWLEDGMENT

This research has been funded by Tactica, COINS (TEC2013-47016-C2-1-R) and EUIN2015-62758 projects, Mineco, Spain, project 24500701 at University of Oulu, and NEWCOM# FP7 project (Grant 318306).